\documentclass[]{aa}

\usepackage{graphicx}
\usepackage{subfigure}
\usepackage{psfrag}
\usepackage{natbib}

\usepackage[T1]{fontenc} % Computer Modern (CM) fonts
\usepackage{ae,aecompl} % and: dvips -Pcmz -Pamz macros_aaa.dvi

\usepackage{verbatim}

\begin{document}

\title{The spectroscopic binary system Gl 375\thanks{The authors are
    visiting astronomers of the Complejo Astron\'omico El Leoncito
    operated under agreement between the Consejo Nacional de
    Investigaciones Cient\'\i ficas y T\'ecnicas de la Rep\'ublica
    Argentina and the National Universities of La Plata, C\'ordoba and
    San Juan.}}  \subtitle{I. Orbital parameters and chromospheric
    activity}

\author{R.F. Díaz\inst{1}, J.F. González\inst{2}, C. Cincunegui\inst{1}, P.J.D. Mauas\inst{1}}

\institute{Instituto de Astronom\'\i a y F\'\i sica del Espacio (IAFE),
             CC. 67, suc. 28, 1428. Buenos Aires, Argentina.\\
             \email{rodrigo@iafe.uba.ar}\and
             Complejo Astronómico El Leoncito (CASLEO), San Juan, Argentina}

\date{}

\abstract {}{We study the spectroscopic binary system Gl~375.} {We
employ medium resolution \emph{echelle} spectra obtained at the 2.15 m
telescope at the Argentinian observatory CASLEO and photometric
observations obtained from the ASAS database.} {We separate the
composite spectra into those corresponding to both components. The
separated spectra allow us to confirm that the spectral types of both
components are similar (dMe3.5) and to obtain precise measurements of
the orbital period (P = 1.87844 days), minimum masses
($M_1\sin^3 i = 0.35\,\mathrm{M}_\odot$ and
$M_2\sin^3 i =0.33\,\mathrm{M}_\odot$) and other orbital
parameters. The photometric observations exhibit a sinusoidal
variation with the same period as the orbital period. We interpret
this as signs of active regions carried along with rotation in a
tidally synchronised system, and study the evolution of the amplitude
of the modulation in longer timescales. Together with the mean
magnitude, the modulation exhibits a roughly cyclic variation with a
period of around 800 days. This periodicity is also found in the flux
of the Ca II K lines of both components, which seem to be in
phase.}{The periodic changes in the three observables are interpreted
as a sign of a stellar activity cycle. Both components appear to be in
phase, which implies that they are magnetically connected. The
measured cycle of $\approx$ 2.2 years ($\approx 800$ days) is consistent
with previous determinations of activity cycles in similar stars.}


\authorrunning{D\'\i az et al.}
\titlerunning{The binary system Gl~375}

\maketitle

\section{Introduction}

\object{Gl~375} (V* LU Vel, HIP 48904) is a $V=11.27$,
$B-V=1.56$~\citep{hipparcos} spectroscopic binary system located 16 pc
away in the Southern constellation Vela, at RA (J2000) = 09:58:34 and
DEC (J2000) = -46:25:30. Both components of the system exhibit large
levels of chromospheric emission.

Until recently, both the duplicity of the system and its high level of
  chromospheric activity were unknown. A considerable number of
  photometric measurements exist in the literature, including very
  early works \citep[see for
  example][]{kron57,corben72,doylebutler90,bessel90}. However,
  the flare activity of Gl~375 was discovered in a relatively recent
  work by \citet{doyle90} using optical spectroscopy and optical and
  infrared photometry. They reported a high rate of flare occurrence
  and classified Gl~375 as dM3.5e, similar to the spectral
  classification of \citet{bidelman85}, who, however, did not indicate
  the presence of emission lines. The spectral resolution of
  \citet{doyle90} was not enough to resolve the double features in the
  spectra of Gl~375, but they noted that the FWHM of the $H_\alpha$
  line was twice that found in the spectra of a similar
  star. Therefore, they concluded that Gl~375 was either a fast
  rotator or a binary. Only recently, \citet{christian2002} obtained a
  high-resolution optical and IR spectrum of Gl~375, which allowed
  them to identify the characteristic double features and report the
  binary nature of this object.

Stars in close binary systems tend to show higher levels of activity
than single stars with the same characteristics \citep[see, for
example][]{schrijverzwaan91,zaqarash2002}. Due to its vicinity and relative
brightness, the present system constitutes an interesting opportunity
to study two very active stars in interaction.

The $\alpha\Omega$ dynamo is accepted to be the cause of activity
cycles in late-type stars with an outer convective layer. This dynamo
is thought to be the result of the action of differential rotation at
the interface between the convective envelope and the radiative
core. Therefore, the presence and characteristics of activity cycles
are closely related to the existence and depth of an outer convection
zone. Since this depth depends on spectral type --from F stars that
have shallow convection zones to middle M stars that are totally
convective--, it is of special interest to study these cycles in stars
of different spectral types, and in particular in middle-M stars, to
determine whether there is an onset of cyclic activity. Recently, an
activity cycle has been detected in Proxima
Centauri~\citep{proxima}. To our knowledge, this is the only fully
convective M-type star for which an activity cycle has been detected
so far. This poses a question on the origin of such a cycle.

The recent discovery of a 5 Earth-mass planet in the habitable zone of
the M3 dwarf \object{Gl~581}~\citep{udry2007} has made the study of
habitability in these stars increasingly relevant~\citep[see,
for example,][]{buccino2006,buccino2007,vonbloh2007}. Changes in the
rate of flare occurrences and CMEs characteristic of activity cycles
could in principle affect the habitability of a planet in close
orbit.

Gl~375 was observed as part of an ongoing program to study
chromospheric activity in late-type main sequence stars. By the time
of our first observations in the year 2000, its binary nature was
unknown. However, it was quickly realized that this was a double
system, with chromospheric emission which was similarly strong in
both components.

In this first paper we present medium-resolution spectra of Gl~375
spanning over six years. We obtain precise measurements of the orbital
parameters of the system and find evidence for the presence of an
activity cycle in both components. In Sect.~2 we describe the
spectroscopic observations and outline the reduction procedure. In
Sect.~3 we describe the method used to separate the spectra of each
component and we present precise measurements of the orbital
parameters of the system. In Sect.~4 we analyze photometric and
spectroscopic data of Gl~375 and find periodic variations consistent
with an acticity cycle. Finally, in Sect.~5 we summarize our results
and outline our conclusions.

\section{Observations and data reduction \label{sec.obs}}
The spectroscopic observations were obtained using the REOSC
spectrograph in the Jorge Sahade 2.15 m telescope at the Complejo
Astron\'omico El Leoncito (CASLEO), located 2550 m above sea level in
the Argentinian Andes. The spectra cover the entire optical range
(from $\approx$ 3800 to 6800 \AA) in 24 \emph{echelle} orders. We
employed a 300 micron-width slit which provided a resolving power of
$R = \lambda/\delta\lambda \approx 13200$. The detector used was a
1024 x 1024 pixel TEK CCD camera.

\begin{table}
\begin{minipage}[t]{\columnwidth}
\caption{Observation log for Gl~375}
\label{log}
%\centering 
\renewcommand{\arraystretch}{1.2}
\renewcommand{\footnoterule}{}
\begin{tabular}{l l l l l}
\hline \hline
Date\footnote{In \emph{ddmmyy} format}&HJD\footnote{HJD-2450000}&Exp Time
&Orbital phase &Notes\footnote{See Sect.~\ref{magact}}\\
&&[sec]&&\\
\hline
020399 & 1241.6469 & 4200 & 0.3572 &{\bf \S}\\
030399 & 1242.5470 & 3000 & 0.8363 &\\
210300 & 1626.5748 & 9600 & 0.2756 &\\
220300 & 1627.5486 & 9540 & 0.7940 &{\bf !}a\\
040301 & 1973.6596 & 14400& 1.0480 &\\
040701 & 2095.4968 & 2400 & 0.9087 &\\
021201 & 2246.7938 & 9600 & 0.4525 &\\
300302 & 2364.5754 & 6000 & 1.1541 &\\
250602 & 2451.4672 & 4800 & 0.4115 &\\
160303 & 2715.6247 & 6000 & 1.0371 &\\
130603 & 2804.4781 & 2400 & 0.3387 &\\
071203 & 2981.8507 & 6000 & 0.7639 &{\bf !}a\\
090304 & 3074.6662 & 6000 & 1.1747 &\\
030604 & 3160.5119 & 6000 & 0.8752 &\\
251104 & 3335.8051 & 1200 & 0.1934 &{\bf !}\\
190305 & 3449.6718 & 6000 & 0.8110 &{\bf !}\\
231105 & 3698.8419 & 7200 & 0.4580 &\\
241105 & 3700.8344 & 6000 & 0.5187 &\\
120206 & 3779.7498 & 9600 & 0.5297 &\\
\hline
\end{tabular}
\end{minipage}
%The first column indicates the
%  date of the observation in \emph{ddmmyy} format. The second column
%  is the heliocentric julian day. The third and fourth columns
%  indicate the exposure time and the orbital phase, respectively. The
%  last column contains notes on the observations (see
%  text).}
\end{table}
Observations began in 1999 and are still ongoing. We obtain about three observations a year, weather permitting. A log of the
  observations analyzed here is presented in Table~\ref{log}. Each log
  entry actually consists of two individual observations which are
  combined to remove cosmic rays, and the reported exposure time is
  the sum of the exposure times of the two individual
  observations. The individual spectra are used in Sect.~\ref{magact}
  to look for flares in the observations.

The observations were reduced and analyzed using standard
    IRAF\footnote{IRAF is distributed by the National Optical
    Astronomy Observatories, which are operated by the Association of
    Universities for Research in Astronomy, Inc., under cooperative
    agreement with the National Science Foundation.} routines. We
    calibrate in flux the spectra, using low-dispersion long-slit
    spectra of the same star. For details of the reduction and flux
    calibration process we refer the reader to \citet{library}.

The resulting spectral resolution is high enough to clearly resolve
the double features in the spectra of Gl~375 at appropriate
phases. The exposure times ranged from 20 minutes to 240 minutes,
producing a wide range of S/N.

Our spectral coverage allows us to study the effect of chromospheric
activity in the whole optical spectrum simultaneously. In previous
works we have studied the behaviour of different lines in our complete
stellar sample (see \citealp{indices} for the H \& K Ca II lines and
H$_\alpha$, \citealp{dobleteD} for the Na I D lines, and
\citealp{magnesio} for the Mg II $h$ and $k$ UV lines).

\begin{figure*}
%\sidecaption
\center
%bb=80 255 540 525 
\includegraphics[width=12cm,clip]{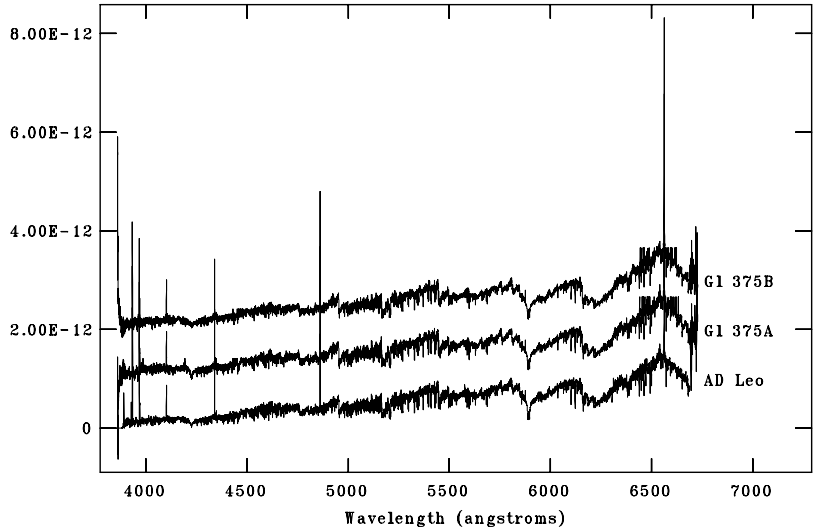}
 \caption{Mean spectra of the two components of the binary system
  Gl~375, together with the spectrum of AD\,Leo used as template. The
  entire spectral range observed is shown, and the spectra are
  displaced arbitrarily for clarity. As can be seen, the three spectra
  are remarkably similar, indicating that both the primary and
  secondary components of Gl~375 have spectral type M3.5Ve.}
  \label{adleo}
\end{figure*}
\begin{figure*}
  \includegraphics[width=0.5\textwidth]{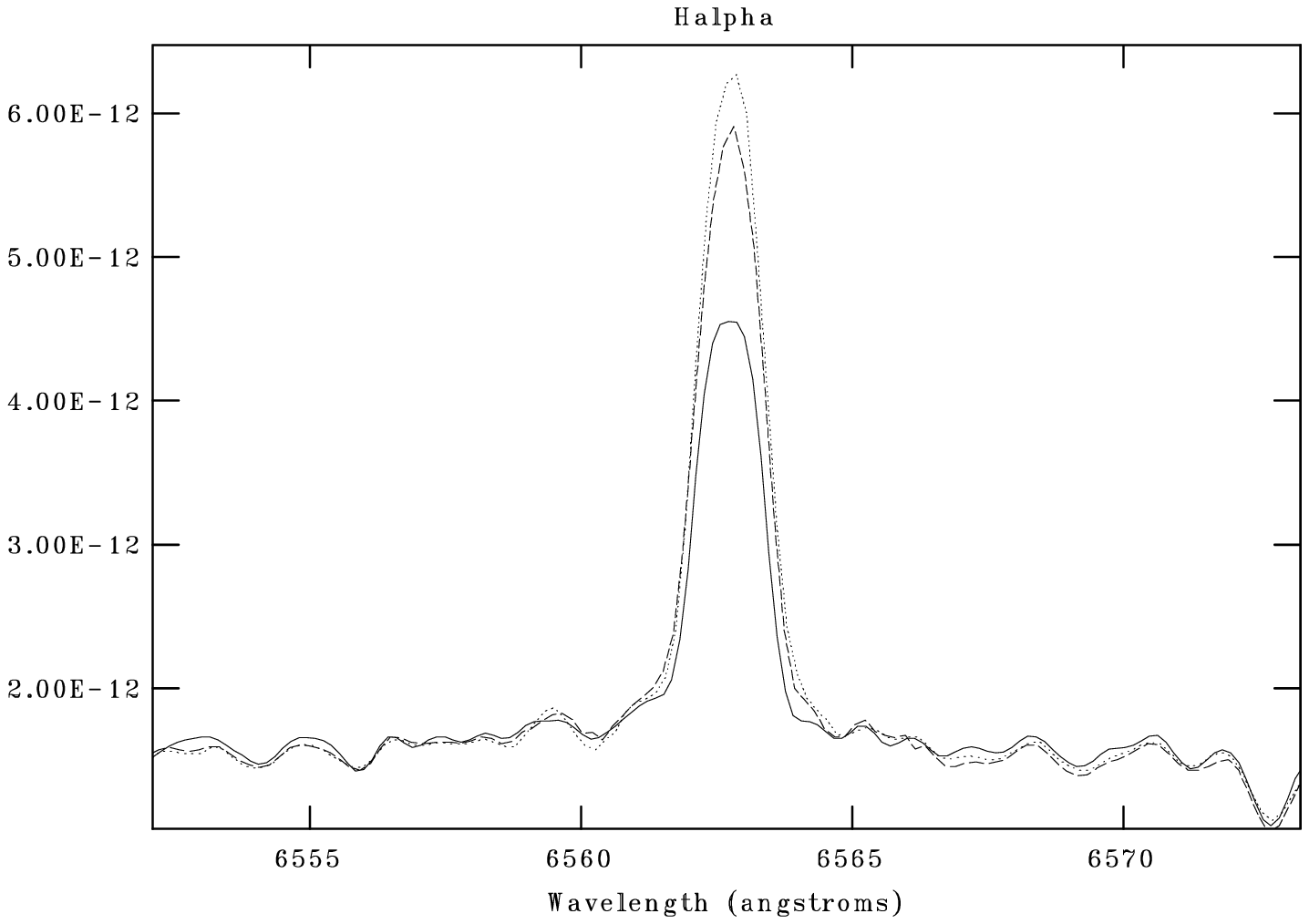}\hfill
  \includegraphics[width=0.5\textwidth]{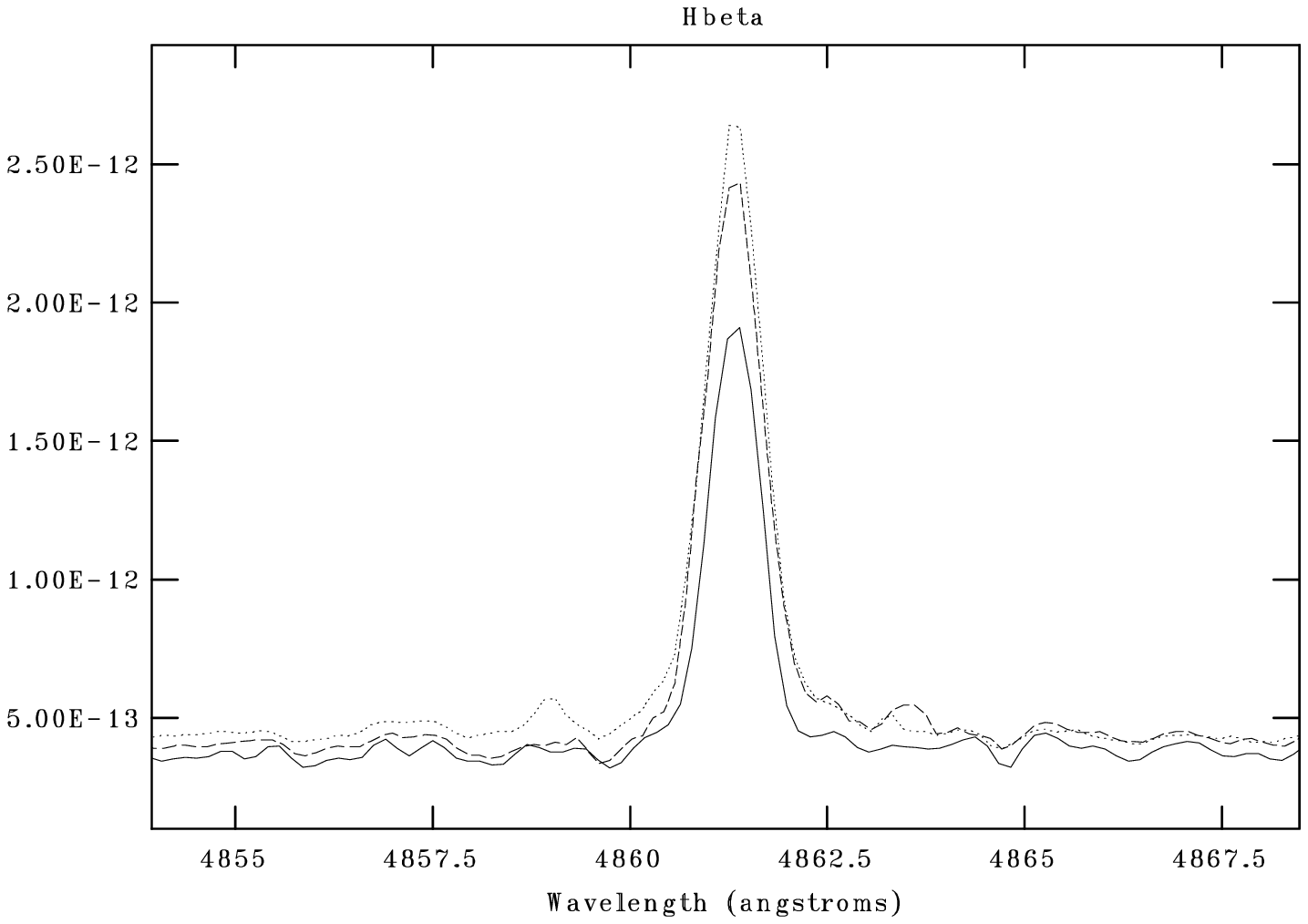}
  \includegraphics[width=0.5\textwidth]{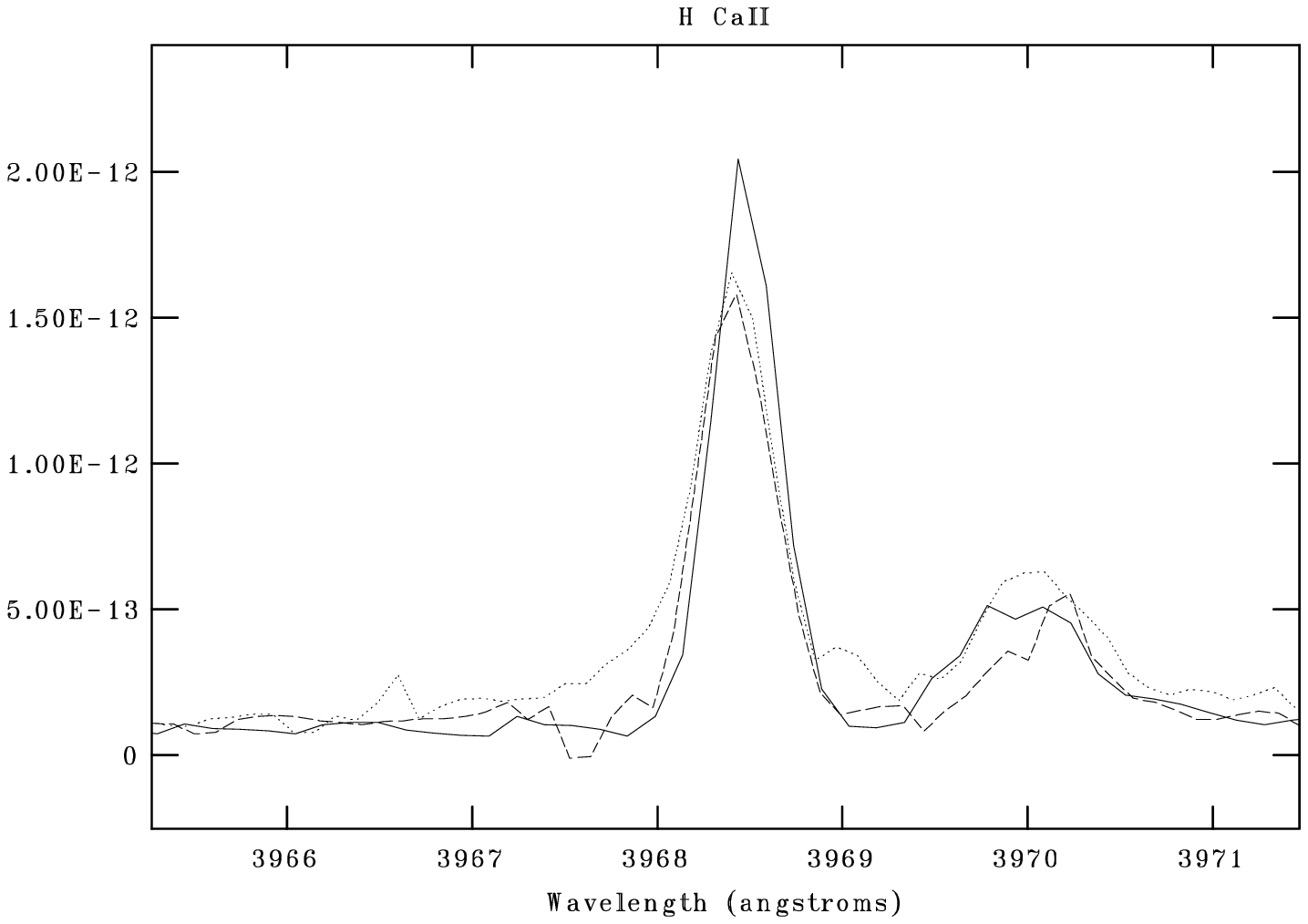}\hfill
  \includegraphics[width=0.5\textwidth]{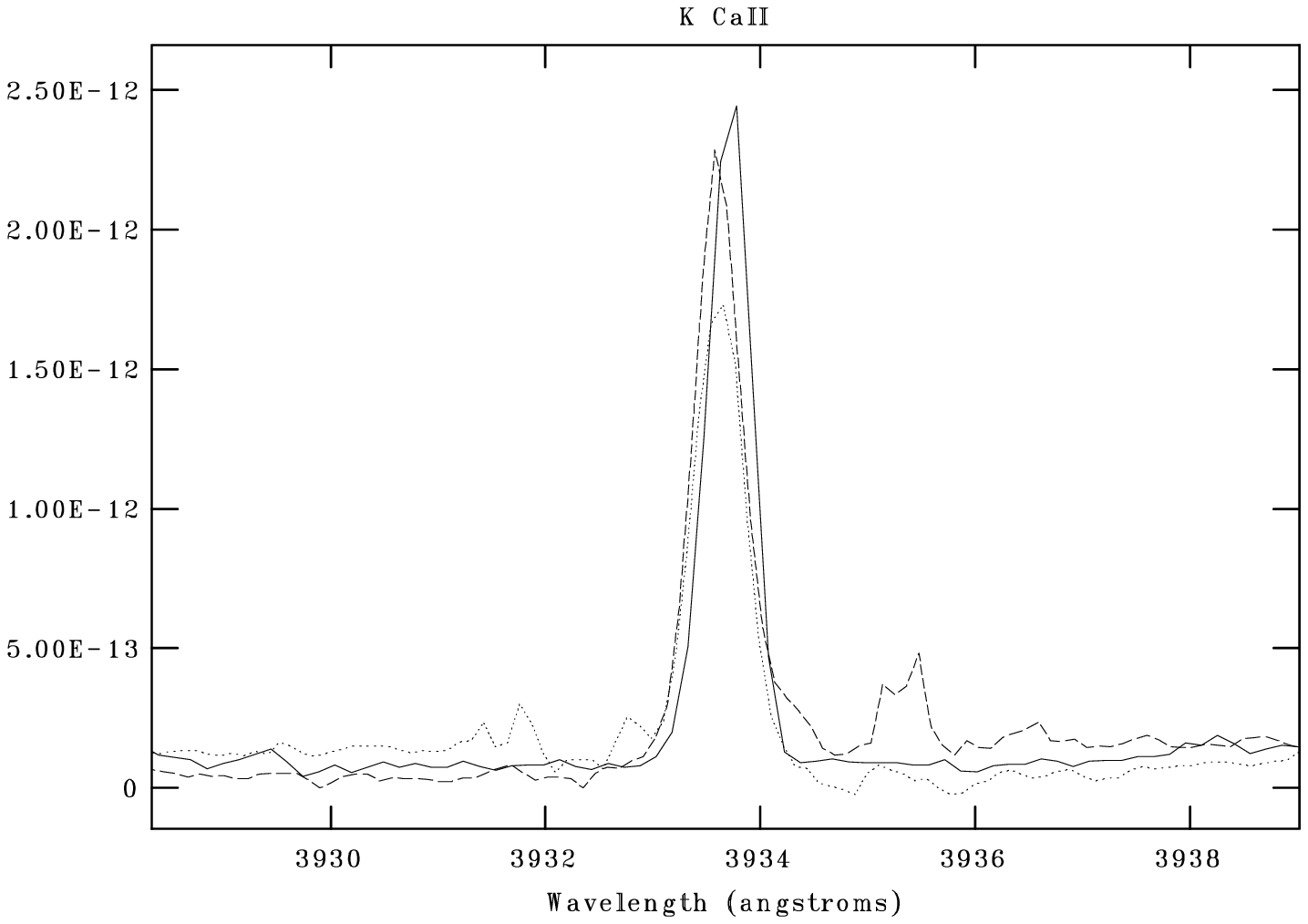}
 \caption{Comparison of the H$\alpha$, H$\beta$, Ca II H and Ca II K
 line emissions in AD\,Leo (solid line), Gl~375A (dashed line) and
 Gl~375B (dotted line). The spectra of Gl~375 are mean spectra
 obtained from the entire set of observations.}
  \label{linecomp}
\end{figure*}
\section{Spectra separation and orbital parameters \label{sec.sep}}
The composite spectra were separated using the iterative method
presented by~\citet{gonzalezlevato2006}. This method allows to compute
the individual spectra and the radial velocities (RVs) of the two
stellar components of the binary system. In each iteration the
computed spectrum of one component is used to remove its spectral
features from the observed spectra. The resulting single-lined spectra
are then used to measure the RV of the remaining component and to
compute its spectrum by combining them appropriately.

\begin{figure*}[!ht]
\center
%\sidecaption
  \includegraphics[width=12cm]{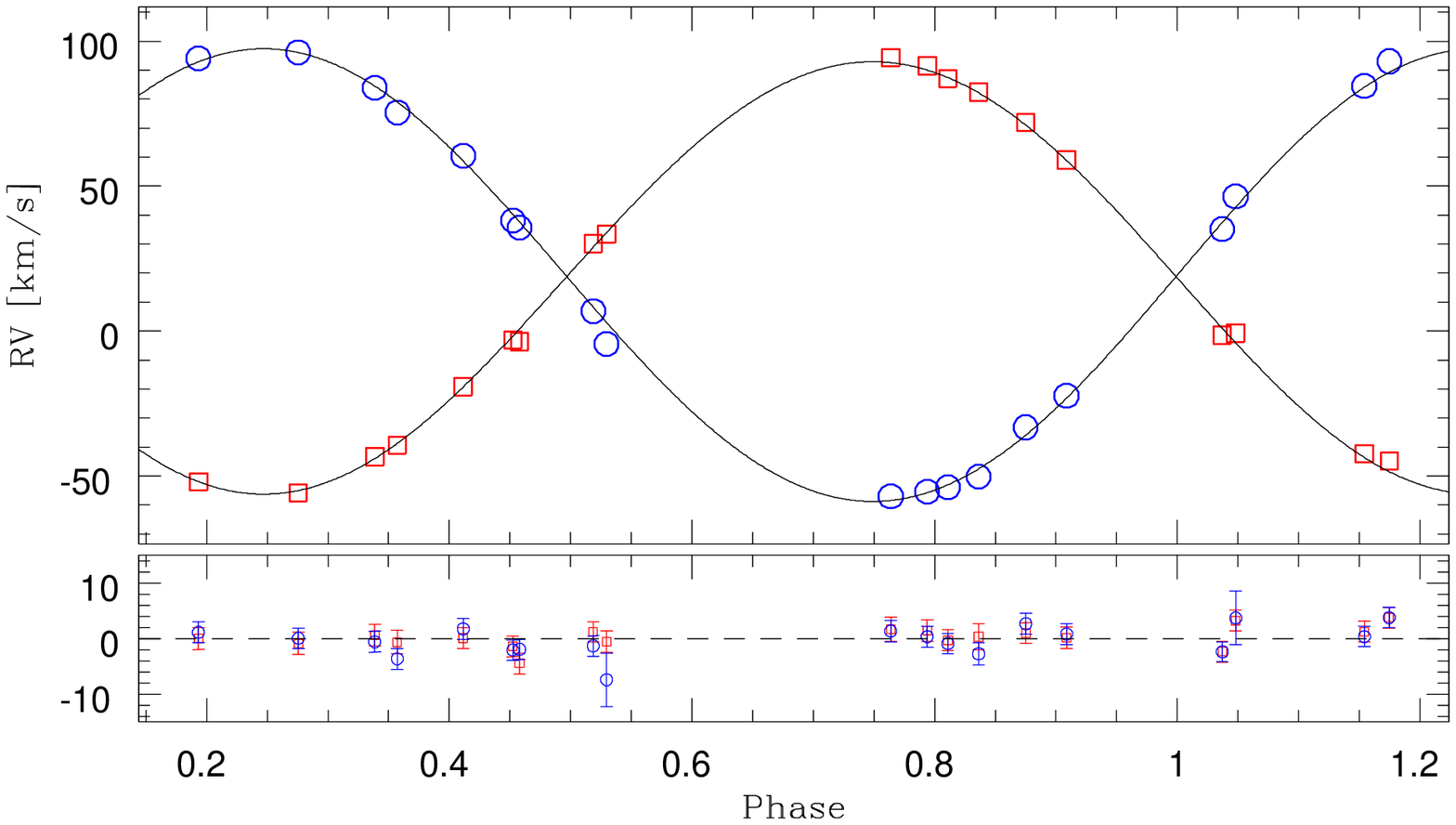}
  \vspace{-0.4cm}\caption{RVs measurements phased to the orbital
  period $P=1.8784425$ days. The squares represent
  measurements of the primary star and the circles are measurements of
  its companion. The solid line is the best fit to the data. In all
  cases the error bars are smaller than the symbols. In the lower
  panel we show the residuals of the fit and the individual errors.}
  \label{orbita}
\end{figure*}

We used as a template for the cross-correlations a spectrum of a star
of similar spectral type (\object{AD\,Leo} = Gl 388) observed with the
same instrumental setup during the night of March 7 2004. The
zero-point was established with an uncertainty of about 0.5
km\,s$^{-1}$ by means of the observation of RV standard stars. In all
cases for RV measurements we excluded the emission lines.

A mean spectra of very high S/N was obtained for each component, which
were used to check the spectral type of the stars.  These spectra are
shown in Fig.~\ref{adleo} together with the spectrum of AD\,Leo we
employed. To measure the spectral type of the components we
employed the TiO bands used by~\citet{pettersenhawley89} to compute
the intrinsic brightness of a sample of red flare
stars. Unfortunately, only three of these bands fall within our
spectral range (4760 \AA, 4950 \AA, 5450 \AA). We carefully compared
the strength of these bands in the mean spectra of each component and
in AD Leo and found that both stars are very similar between them, and
similar to AD Leo, one of the dM3.5 stars used
by~\citet{pettersenhawley89}. Therefore, the spectral types of the
components must be around M3.5. Additionally, we computed the absolute
magnitude of each component using the parallax from the Hipparcos
catalogue~\citep{hipparcos} and our result that both stars are
identical. We found that both components have absolute magnitude $M_v
= 11.01$, which is in excellent agreement with the value for AD Leo
$M_v = 10.98$ \citep[see][]{johnson65,vanaltena95}. Unfortunately, we
could not verify the obtained spectral type directly by using the
relations given by~\citet{reid95} and \citet{kirkpatrick91} since the
TiO bands employed in those papers lay outside our spectral range.

Also evident from Fig.~\ref{adleo} is the high level of emission in the
Balmer lines and in the H \& K Ca II lines, which implies a high level
of chromospheric activity.  In Fig.~\ref{linecomp} we compare the flux
intensity of AD Leo and Gl~375 in H$_\alpha$, H$_\beta$, and in the Ca
II H \& K lines. Note that the line intensities in the mean spectra of
Gl~375A and Gl~375B are comparable to those from AD Leo, which is
considered a very active flare star. This should be expected since, as
mentioned in the Introduction, stars in close binary systems usually
exhibit enhanced levels of chromospheric activity when compared with
single stars of the same
characteristics~\citep{schrijverzwaan91,zaqarash2002}. In
Sect.~\ref{magact} we present evidence of a continuous interaction
between the magnetospheres of the components which may lead to an
enhanced level of chromospheric activity.

Using the obtained values of RV we fit a Keplerian orbit to the
system. In Fig.~\ref{orbita} we show the RV curves along with the
resulting fit. %In the lower panel we present the residuals of the fit
%and the errors. In the upper panel the error bars are in all cases
%smaller than the symbol size.
The orbital parameters we obtained are presented in
Table~\ref{parametros}. Note that since the minimum mass of the
components is close to the expected value for M-type dwarves, which is
around $0.4\,\mathrm{M}_\odot$ for spectral type M2 and
$0.22\,\mathrm{M}_\odot$ for M5 \citep{allen}, the orbital inclination
angle cannot differ much from 90$^\circ$. Therefore the minimum
distance $a\sin{i} = (5.665\pm0.035)\,\mathrm{R_\odot}$ should be
similar to the actual separation between the components. For such a
tight orbit, we would expect to find the stars tidally locked. In this
respect, \citet{zahn77} studied the effects of turbulent viscosity in
binary star systems and provided expressions for the characteristic
time for orbital and rotational synchronization and orbital
circularization. Employing Eqs. 6.1 and 6.2 from that work, we find
that for a system as the one studied here, the synchronisation time is
\begin{displaymath}
t_\mathrm{sync} \approx (q)^{-2}\;(a/R_1)^6 \approx 8.9\,\mathrm{Myr}
\end{displaymath}
and the circularization time is
\begin{displaymath}
t_\mathrm{circ} \approx (q(1+q)/2)^{-1}\;(a/R_1)^8 \approx 1.7\,\mathrm{Gyr} \; \; ,
\end{displaymath}
where $q = M_2/M_1$ and $R_1$ is the radius of the primary. For these
calculations we have assumed the primary radius to be $0.4\,R_\odot$
\citep[see][]{allen} and the orbital separation to be similar to its
minimum value, $a \approx 5.66\,R_\odot$. Therefore, since we found a value of the
eccentricity which is undistinguishable from zero, we believe the
system must have its orbital and rotational periods
synchronised. However, to our knowledge no measurements of the period
of rotation of this system exist in the literature. In
Sect.~\ref{sec.rot} we provide further evidence that the rotational
and orbital periods are indeed synchronised.

%\citet{drake98} studied the synchronization timescales in stars
%with Jupiter-mass companions. They compare the timescales of
%synchronization by means of turbulent viscosity
%\citep{zahn77,zahn89,rieutordzahn97} with those arising from
%considering hydrodynamic spin-down \citep{tassoul87, tassoultassoul90,
%tassoultassoul92}. In both cases, the characteristic timescales for
%synchronization in our system is

\begin{table}[ht]
\begin{minipage}[t]{\columnwidth}
\caption{\label{parametros} Orbital parameters for Gl 375.}
\renewcommand{\arraystretch}{1.2}
\renewcommand{\footnoterule}{}
\renewcommand\thefootnote{\it{\alph{footnote}}}
%\scriptsize
\begin{tabular}{l l}
\hline \hline
Conjunction epoch [HJD]&$2452779.422\pm0.005$\\
Period [days]&$1.87844246\pm 0.0000049$\\
$a\sin{i}$ [$\mathrm{R_\odot}$]&$5.665\pm0.035$\\
Systemic velocity [$\mathrm{km\,s^{-1}}$]&$18.81\pm0.32$\\
Primary RV amplitude [$\mathrm{km\,s^{-1}}$]&$74.58\pm0.68$\\
Secondary RV amplitude [$\mathrm{km\,s^{-1}}$]&$78.07\pm0.66$\\
Orbital excentricity&$0.0072\pm0.0070$\\
$M_1\sin^3{i}$ [$\mathrm{M_\odot}$]&$0.3541\pm0.0068$\\
$M_1$[$\mathrm{M_\odot}$]&$> 0.3646$ \footnote{Infered from the absence of eclipses in the lightcurve}\\
$M_2\sin^3{i}$ [$\mathrm{M_\odot}$]&$0.3382\pm0.0068$\\
$M_2$ [$\mathrm{M_\odot}$]&$> 0.3482$ \footnotemark[1]\\
\hline
\end{tabular}
\end{minipage}
\end{table}

\section{Variability}
In this Section, we study the variability of Gl~375 as a whole and of
each component. Besides using the spectra obtained at CASLEO, we also
employed photometric observations provided by the All Sky Automated
Survey~\citep[ASAS,][]{pojmanski2002}. The ASAS data for this system
cover the period between 2002 and 2006, with gaps during the year 2002
and 2005. We discarded all the observations which were not qualified
as either A or B in the ASAS database (i.e. we retained only the best
quality data), as well as 7 outlier observations. The final dataset
consists of 295 points, with typical errors of around 15 mmag.

\subsection{Mean Magnitude \label{sec.mag}}
\begin{figure}
 \centering
  \includegraphics[width=\columnwidth]{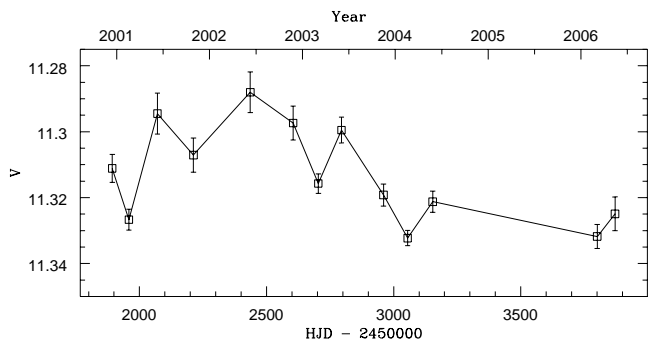}
  \caption{Mean V magnitude of GL 375 as measured by the All Sky
  Automated Survey as a function of time. The mean was taken every
  four months in each observing season.}
  \label{vmag}
\end{figure}

In Fig.~\ref{vmag} we plot the mean V magnitude of the system as a
function of time, as measured by ASAS. The empty squares
represent the weighted mean of the observations, taken over four
months in each observing season and the error bars represent the
square root of the variance of the weighted mean, computed as
\citep[see][Eq. 9.12]{frodesen}:
\begin{displaymath}
Var = \left(\sum^n_{i=1} 1/\sigma_i^2\right)^{-1} \; \; ,
\end{displaymath}
where $\sigma_i$ is the photometric error of $i$th point, which is
provided by ASAS, and $n$ is the total number of observations
analized. Note the lack of observations during part of the 2002
season, as well as in the entire 2005 and the beginning of 2006
seasons.

A maximum is discernable around the year 2002.5 ($HJD\approx
 2452430$), and a minimum in 2004 ($HJD\approx 2453050$). A
few additional minima seem to be present near 2002 ($HJD\approx
2452200$) and 2003.25 ($HJD\approx 2452700$). Unfortunately, the lack
of observations during the last part of 2004 and 2005 precludes any
further analysis at this point, since a maximum could have occurred in
2005. We return to this issue in Sect.~\ref{magact}.

\subsection{Rotational Modulation \label{sec.rot}}
\begin{figure}
 \centering \includegraphics[width=\columnwidth]{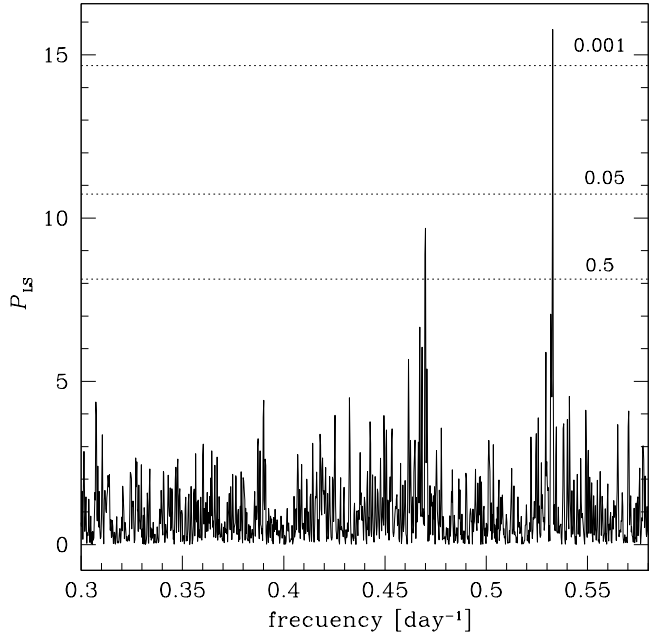}
  \caption{Lomb-Scargle periodogram of the ASAS data. The main peak occurs
  at $\mathrm{P_{phot}} = 1.876644$ days, and has a False Alarm
  Probability of about $0.1\%$. The horizontal dashed lines show
  the values of the False Alarm Probability.}
  \label{periodogram}
\end{figure}

We calculated the Lomb-Scargle periodogram
\citep{scargle82,hornebaliunas86} for the ASAS data. This
periodogram is a method to estimate the power spectrum (in the
frequency domain), when observation times are unevenly spaced, and
normalizes the spectrum in such a way that it is possible to estimate
the significance of the peaks. The amplitude of the periodogram at
each frequency point is identical to the equation that would be
obtained estimating the harmonic content of a data set, at a given
frequency, by linear least-squares fitting to an harmonic function of
time~\citep{press92}. We show this periodogram in
Fig~\ref{periodogram}.

As can be seen, there is a distinct peak corresponding to a period of
$P_\mathrm{phot} = 1.876644$ days. For this peak, the False Alarm
Probability is about $0.1\%$. The excellent False Alarm Probability
indicates the presence of a harmonic variability within the data set
with a period resembling very closely the measured orbital period.

We believe this harmonic variability is produced by spots and
active regions in the stellar surface carried along with
rotation. This would imply that the rotational and orbital periods are
synchronised, as is expected for such a close binary.

\begin{figure*}[!th]
  \includegraphics[width=\columnwidth]{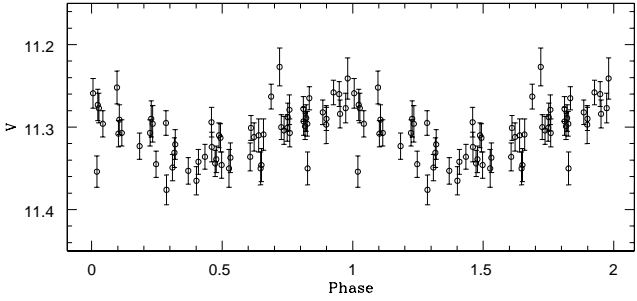}\hfill
  \includegraphics[width=\columnwidth]{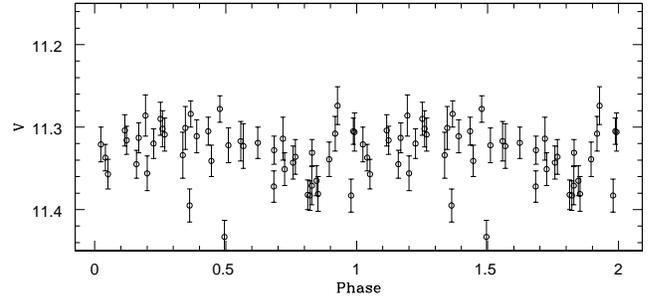}
  \caption{ASAS photometry phased to the orbital period for two
  different epochs. \emph{Left}: 2002.5 - 2003.5 \emph{Right}:
  2006}
  \label{ASAS}
\end{figure*}

To verify if this is indeed the case, we phased the data to the
obtained period, for two different seasons (Fig.~\ref{ASAS}).  The
sinusoidal shape of the variation is evident, although the amplitude
of the modulation is different in both cases: between 2002.5 and
2003.5 (upper panel) the amplitude is about 30 mmag, and falls to
about 15 mmag during 2006. This could indicate a different area
covered by starspots or active regions, and therefore different
activity levels, in each epoch. Note that during the period of minimum
modulation the mean brightness is smaller by about 20 mmag.

Note also that whether the period of maximum rotational modulation
corresponds to an epoch of maximum or minimum activity depends on the
filling factor. On the other hand, the variations in the mean
brightness can be dominated by the active regions (as in the Sun) or
by the spots (as in younger stars), which would cause the stars to be
either brighter or fainter during periods of maximum activity,
respectively. Therefore, considering that during the period of minimum
modulation the obseved system is fainter (see Fig.~\ref{ASAS}), we are left
with two possibilities. Either the system is dominated by active
regions and the filling factor during maximum activity is below 50\%
(as in the Sun), or is the spots that control the photometric
variations and the maximum filling factor is over 50\%.

We found no evidence of eclipses in the photometry, and since the ASAS
data have complete phase coverage in many periods, this could not be
an eclipsing system, and we can therefore set an upper limit to the
inclination angle.  Assuming a stellar radius of
$0.4\,\mathrm{R_\odot}$ for both components, the inclination angle
should be larger than $82^\circ$. In turn, this implies lower limits
for the stellar masses, $M_1>0.3646\,\mathrm{M_\odot}$ and
$M_2>0.3482\,\mathrm{M_\odot}$. Note that the assumed spectral type of
the stars remain consistent with these values.

\begin{figure}
 \centering
  \includegraphics[width=\columnwidth]{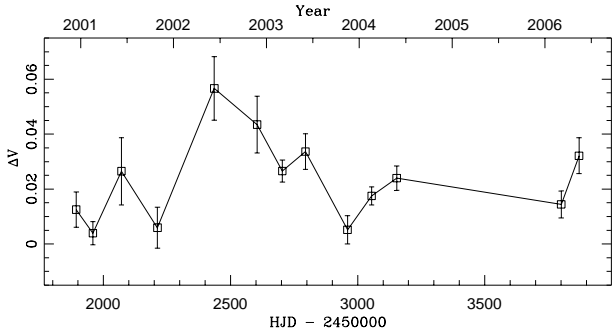}
  \caption{Variation in the binned photometric data as measured from a
  sinusoidal fit (see text) as a function of time. A peak is clearly
  visible near the year 2003.}
  \label{deltamag}
\end{figure}

To study the evolution of the activity level, we studied in greater
detail the changes in the amplitude of the photometric modulation as a
function of time. Assuming that the modulation is produced by active
regions and spots in the stellar surface, its amplitude should trace
the amount of inhomogeneities in the surface, and therefore the level
of magnetic activity.

We binned the ASAS measurements every four months and phased the data
on each bin to the orbital period. Bins in which less than 7
observations were present were discarded. The phased data was then
fitted to a sinusoidal model of the form:
\begin{equation}
a_0 + a_s \sin(2\pi\varphi) + a_c \cos(2\pi\varphi)\; \; ,
\end{equation}
where $\varphi$ is the orbital phase. The resulting fit parameters and
their covariance matrix were employed to measure the amplitude $A$ of
the variation and its standard error $\sigma_A$ in every time bin:
\begin{eqnarray}
A & = & \sqrt{{a_s}^2 + {a_c}^2}\\
\sigma_A & = &A^{-1} [\mathrm{var}(a_s)\,{a_s}^2 + \mathrm{var}(a_c)\,{a_c}^2 +
    2\,a_s\,a_c \cdot \mathrm{cov}]^{1/2}\; \; ,
\end{eqnarray}
where var($a_x$) denotes the variance of the parameter $a_x$ and cov
is the covariance of the parameters. The results are shown in
Figure~\ref{deltamag}, where the time assigned to each measurement was
the mean time of the photometric data considered.

It is interesting to compare the present Figure with Fig.~\ref{vmag}.
  The rotational modulation also exhibits a maximum around the year
  2002.5, as in the previous section, and there are also two marked
  minima around the year 2002 and 2004, although the minima of
  amplitude of the rotational modulation are more significant.
  Consider that although these measurements originate in the same
  dataset, the presence of minima and maxima at the same time is
  meaningful, since the measurements are of different nature.

\begin{figure}[!ht]
  \centering
  \includegraphics[width=\columnwidth]{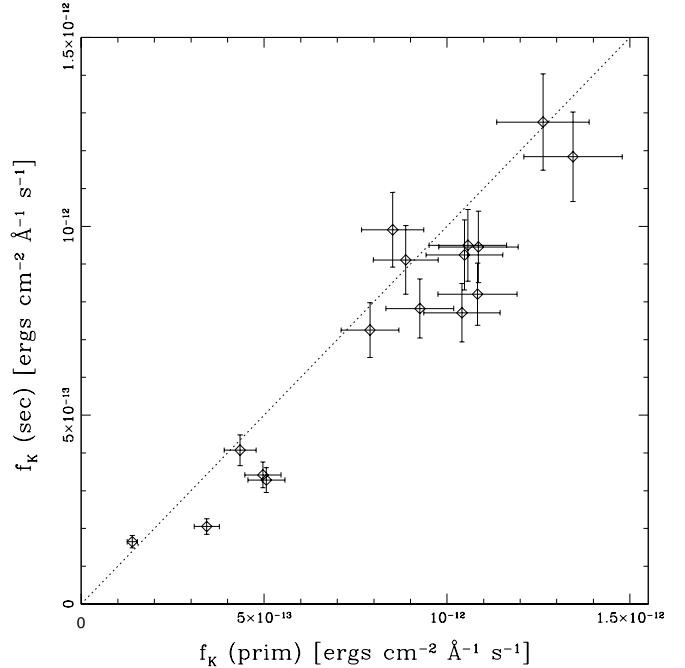}
  \caption{Comparison of the fluxes in the CaII K line for both
  components. The error bars correspond to a 10\% error in the line
  fluxes, and the dotted line is the identity relation.}
  \label{fhvsfh}
\end{figure}
\begin{figure}[!ht]
  \centering
  \includegraphics[width=\columnwidth]{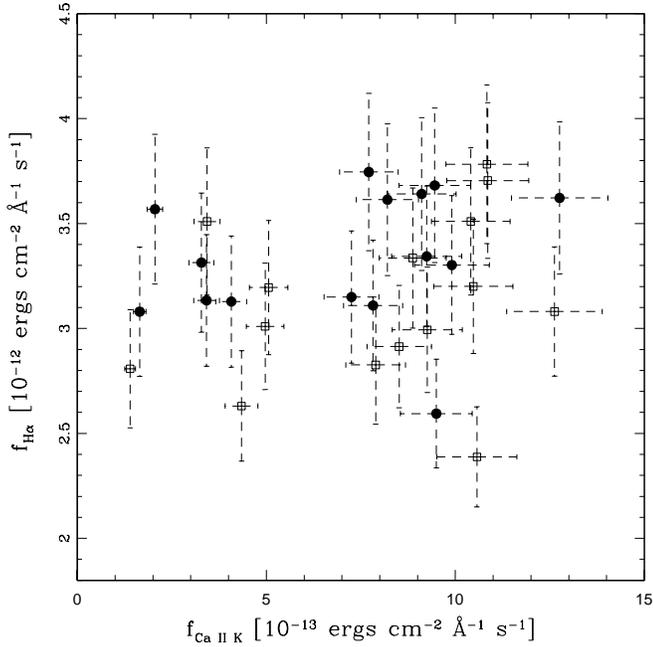}
  \caption{Line fluxes in H$\alpha$ and in the Ca II K line for the
  primary (empty squares) and the secondary (filled circles)
  components of the system. It can be seen that the line fluxes are
  uncorrelated for both stars. The error bars correspond to a 10\%
  error in the line fluxes.}
  \label{fhvsfk}
\end{figure}

\subsection{Magnetic activity \label{magact}}
To study the magnetic activity of the system we use the flux in the
CaII K line, a well-known activity indicator. We did not use the CaII
H line, since it is contaminated by the nearby Balmer line
H$\epsilon$, which we expect to be in emission in these active
stars.

Since we have separated the spectra from both components, we should be
able to distinguish whether the changes observed in the photometric
measurements belong to one of the components or to both. The
spectroscopic measurements also provide further information about the
active regions present in the stellar surfaces. For example, combining
the photometric data with measurements of activity proxies it should
be possible to discern whether active regions or stellar spots
dominate the photometric variations, and whether the filling factor is
larger or smaller than 50\%. Furthermore, since we have observations
for the year 2005, a further detailed study of the system variability
is possible.

We carefully examined the separate spectra and discarded those where
 reduction problems were present or where the S/N was too poor. We
 also discarded spectra which exhibited signs of a transient event,
 such as a flare. To do this we employed the individual observations
 (see Sect.~\ref{sec.obs}) and compared the amplitudes of the K line
 fluxes in one individual spectrum with those in the other. If no
 flare were present, no significant difference should be observed,
 since the time between observations is about an hour. On the other
 hand, if a strong transient event is present during our observations,
 we should observe differences in the K line fluxes from one spectrum
 to the next. In the last column of Table~\ref{log} the discarded
 observations are marked with a {\bf !}  sign. In some cases,
 generally due to changing weather conditions or the presence of
 flares, only one of the individual spectra had to be discarded. In
 those cases, a letter next to the sign indicates which spectra was
 discarded, either a or b, for the first and second one,
 respectively. Additionally, the spectra taken during the night
 of March 2nd 1999 (marked \S) were discarded because the spectral
 range does not include the Ca II K line, due to the use of a
 different instrumental setup during that night. The flux
 measurements were carried out in the remaining individual spectrum,
 after checking that the regions of interest were free of cosmic
 rays. Unfortunately, one of the discarded spectra belong to the year
 2005 and another one to the end of 2004.

\begin{table}
\renewcommand{\arraystretch}{1.2}
\caption{Flux in the Ca II K line for both components of the
  system}
\begin{tabular}{c c c c}
\hline \hline
Name&HJD-2450000&f (primary)&f (secondary)\\
&&\multicolumn{2}{c}{[10$^{-13}$ ergs cm$^{-2}$\AA$^{-1}$s$^{-1}$]}\\
\hline
030399 & 1242.5470 & 10.48 & 9.24\\
210300 & 1626.5748 & 10.41 & 7.71\\
220300 & 1627.5486 & 12.62 &12.76\\
040301 & 1973.6596 & 10.57 & 9.50\\
040701 & 2095.4968 &  4.35 & 4.07\\
021201 & 2246.7938 &  7.89 & 7.25\\
300302 & 2364.5754 & 10.86 & 9.46\\
250602 & 2451.4672 &  8.51 & 9.91\\
160303 & 2715.6247 & 10.83 & 8.20\\
130603 & 2804.4781 &  5.06 & 3.28\\
071203 & 2981.8507 & 13.44 &11.84\\
090304 & 3074.6662 &  9.26 & 7.82\\
030604 & 3160.5119 &  8.87 & 9.11\\
231105 & 3698.8419 &  1.40 & 1.65\\
241105 & 3700.8344 &  4.97 & 3.42\\
120206 & 3779.7498 &  3.43 & 2.05\\
\hline
\end{tabular}
\label{fluxes}
\end{table}
The fluxes in the K line were computed in a 1 \AA\ window using the
 IRAF task SBANDS, which measures the flux by summing each pixel in
 the defined interval. At the edges of the bandpass it considers only
 the fraction of the pixel laying within the bandpass. The results are
 presented in Table~\ref{fluxes}. In Figure~\ref{fhvsfh} we plot the K
 line flux for both components of the system. The error in the line
 fluxes was conservatively estimated as 10\%. It can be seen that the
 fluxes are strongly correlated ($r=0.95$) and, therefore, the
 activity variations of the components should be in phase. For this to
 occur, the magnetospheres of the stars must be strongly
 interacting. \citet{vahia95} simulated the interaction of the
 magnetic field of binary stars. In that work, the stars are modelled
 as perfect dipoles surrounded by vacuum and it is shown that it
 should be quite common to find one of the components located
 completely inside the magnetosphere of the other. Various works
 describe mechanisms by which magnetic activity can be enhanced in
 interacting systems like this one \citep[see, for example,
 ][]{zaqarash2002}. This kind of mechanisms could in principle explain
 the high level of emission present in the lines used as activity
 proxies, such as the Ca II K line. In Figure~\ref{fhvsfh} the dotted
 line is the identity relation. Note that the line fluxes of the
 primary component are systematically larger than those of the
 secondary. The mean flux of the primary is around 13\% larger than on
 the secondary, and the difference appear to be greater for larger
 fluxes.

\begin{figure}
  \centering
  \includegraphics[width=\columnwidth]{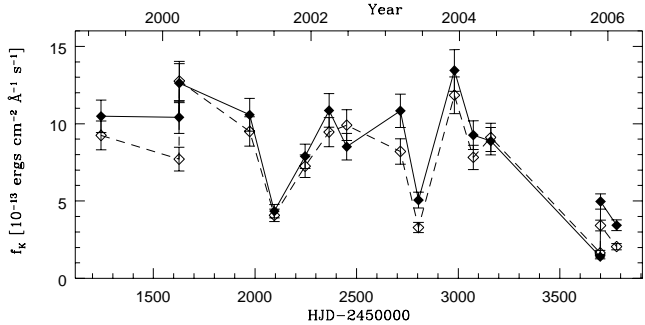}
  \caption{Flux in the CaII K line as a function of time for
  the primary (solid line) and secondary (dashed line) components of
  the system. The error bars correspond to an 10\% error in the line
  fluxes.}
  \label{fhvsjd}
\end{figure}
In Fig.~\ref{fhvsjd} we plot the flux in the K line as a function of
time. The filled (empty) diamonds and solid (dashed) lines represent
observations of the primary (secondary) component. Distinct
maxima are present for both components around the years 2002 and
2004. There are also significant minima at 2001.5 and 2003.5, and a
less clear one at around 2006. Apparently, the periods with little
activity (minima) are much shorter than those with higher level of
chromospheric emission, and are separated by about 2 years.

Another activity indicator usually employed in the study of M stars is
the flux in H$\alpha$. The flux in this line is generally considered
to correlate well with the flux in the CaII H and K lines and has the
advantage of being located in a redder wavelength range than the Ca II
lines, where M dwarfs are brighter. However, it has been recently
reported by~\citet{indices} that the correlation between Ca II and
H$\alpha$ is not always valid. In that paper we found that while
some stars exhibit correlations between H$\alpha$ and the Ca II lines,
the slopes change from star to star. Furthermore, other stars exhibit
anticorrelations and there are also cases where no correlation is
present. Therefore, we checked whether a correlation existed for the
components of Gl 375. In Fig.~\ref{fhvsfk} we plot the lines fluxes in
H$\alpha$ vs. those in the Ca II K line for both components of Gl 375.  It
can be seen that within the errors, none of the components exhibit any
sign of correlation between the lines fluxes and, therefore, this line
cannot be used as an activity indicator for these stars.

In the next Section we study in detail if the system presents a definite
activity cycle, like the one present in the Sun. To do this, we
compare the three observed magnitudes.

\subsection{Periodicity}
\begin{figure*}
 % \sidecaption
  \center
  \includegraphics[width=\textwidth]{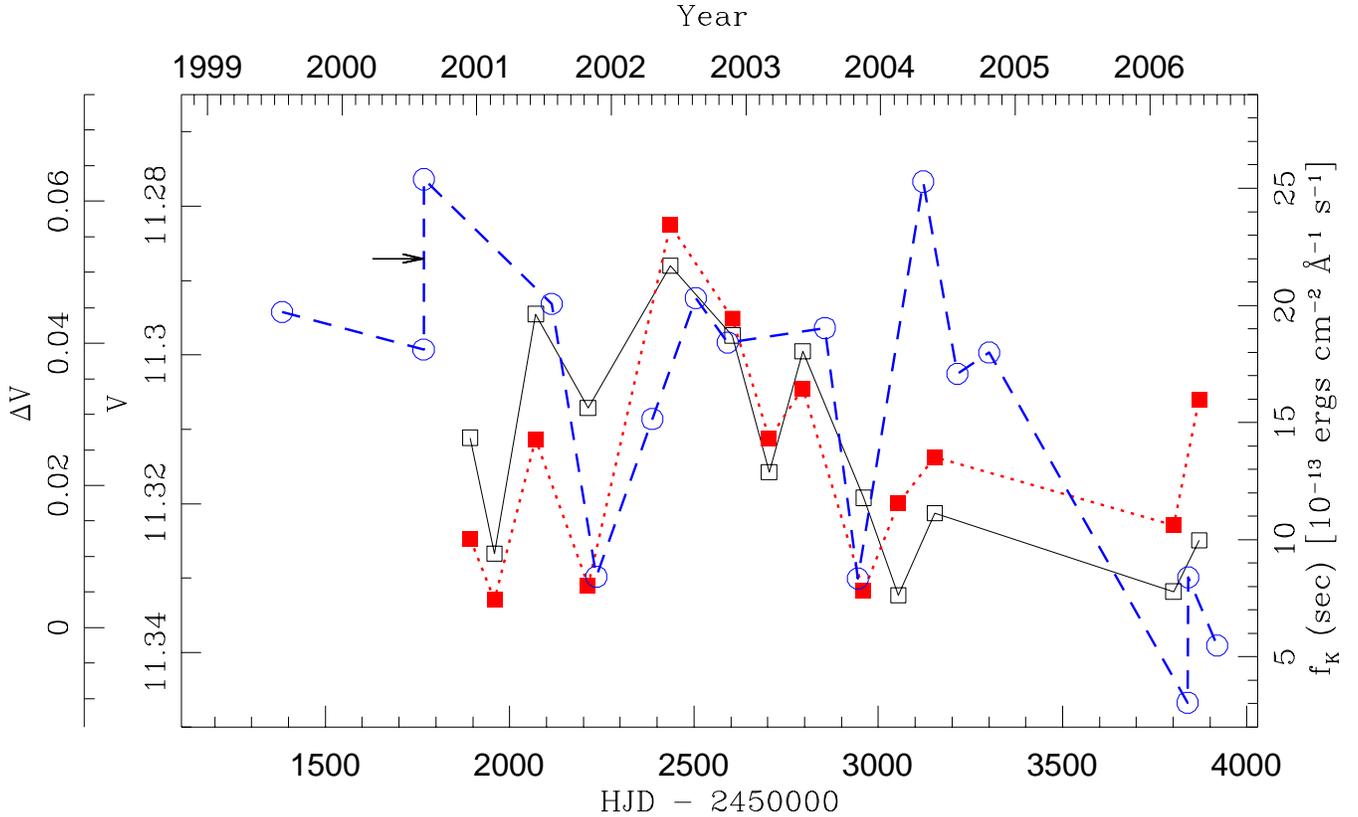}
  \caption{Mean magnitude (black solid line and empty squares),
  amplitude of the rotational modulation (red dotted line and filled
  squares) and flux in the CaII K line (blue dashed line and empty
  circles) as a function of time. The arrow indicates the temporal
  displacement of 140 days that have been applied to the line flux
  (see text).}
  \label{timelag}
\end{figure*}

In Fig.~\ref{timelag} we plot together the three measured quantities:
mean magnitude (empty squares and solid line), amplitude of the
rotational modulation (filled square and dotted line), and the K line
flux (empty circles and dashed line). Since the fluxes in the K lines
of both components are in phase, for clarity in the graph we plot the
total flux of the system. Also to avoid crowding the
graph, we have not included the error bars. The K line flux curve has
been shifted by 140 days to coincide with the other curves. The arrow
in Fig.~\ref{timelag} indicates this displacement.

This timelag between photometric and magnetic variations has already
been observed for stars of different spectral types, from
$\beta\,\mathrm{Com}$ (GOV) to $\epsilon\,\mathrm{Eri}$ (K2V), including
binary systems like $\xi\,\mathrm{Boo}$ \citep[see, for
example,][]{graybaliunas95,gray96a,gray96b}. In particular,
\citet{gray96b} showed that the time lag between magnetic variations,
and variations observed in photometry and temperature is
anticorrelated with effective temperature (see their figure 8). For
the coolest star in their sample ($\epsilon\,\mathrm{Eri}$) the time
lag of temperature variations is about 0.3 years. The timelag present
in the Gl~375 system is comparable to this value.

As mentioned before, the position of the maxima and minima of the mean
magnitude (empty squares and solid line) and those observed in the
amplitude of the rotational modulation (filled square and dotted line)
coincide very well. It can also be seen that, once displaced, the
behaviour of the K line flux also agrees very well with the other two
observables. The maximum around the year 2002.5 is present for all
curves, as well as the minima near 2001.8 and 2004. Note also that the
distinct minimum in the line flux present around the year 2006.2 is
also present in the curves of the other two observables, although the
minima are much less significant, and that the slight decrease in
magnitude and amplitude around 2003.2 coincides with a decrease in the
line flux around 2002.8. Unfortunately, no spectrum was obtained
between the one at 2002.8 and the one at 2003.5.

The agreement between the behaviour of the three observables is
remarkable, because of the different nature of the observations, and
the different instruments and sites in which they were
obtained. Although no clear periodicity can be concluded from the
present data set, the main minima present in all three curves appear
to occur periodically every 2.2 years. This hints to the presence of a
periodic activity cycle in this system, although further observations
are necessary.

\begin{figure}
  \centering
  \includegraphics[width=\columnwidth]{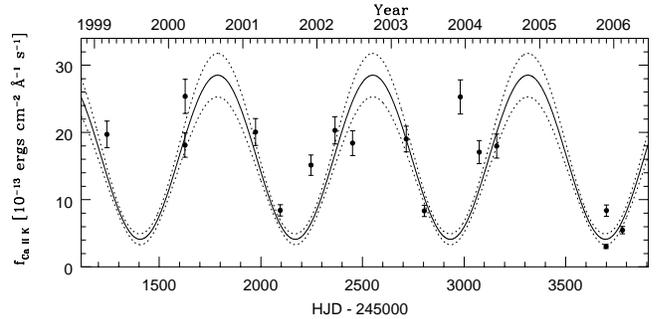}
  \caption{Total flux of the system in the CaII K line as
  a function of time. The error bars correspond to a 10\%
  error in the line fluxes, and the solid and dashed lines are the
  best sinusoidal fit and the $\pm 3 \sigma$ deviations,
  respectively. The period obtained from the fit is 763 days.}
  \label{sinus}
\end{figure}

To further study this point we fit the period, amplitude and
  phase of a sinusoidal function to the total flux using a
  modification of the Levenberg-Marquardt algorithm available in the
  SciPy library for Python. The fit is shown in Fig.~\ref{sinus}
  together with the $\pm 3\,\sigma$ deviations, for the conservative
  estimate of 10\% in the line fluxes. The obtained period is $P =
  763$ days, with a formal error of less than 5 days. However, the fit
  is extremely sensitive to some of the points. For example, if the
  lowest point, located near $\mathrm{HJD} - 2450000 = 3700$ is not
  considered in the fit, the obtained period changes to 786
  days. Therefore, we consider that a more realistic error for the period
  would be of 5\%, i.e. about 40 days.


\section{Conclusions}
We studied the spectroscopic binary system Gl~375 using medium
resolution \emph{echelle} spectra obtained at the Argentinian
observatory CASLEO since the year 2000, and photometric measurements in
the V band from the ASAS database. We found the system exhibits
enhanced levels of chromospheric emission, as is expected for
binary systems.

From each observed spectrum we obtained the individual spectra and RV
of each component. We found that the two components are dM3.5 stars,
and that they orbit each other every $\sim 1.88$ days in a circular
orbit, with a minimum radius $a\sin{i} \sim
5.7\;\mathrm{R_\odot}$. The minimum masses obtained (see
Table~\ref{parametros}) are similar to the expected values for this
type of stars and, therefore, the orbital inclination angle is
expected to be $\sim 90^\circ$. However, the system does not exhibit
eclipses and, therefore, we are able to set an upper limit to the
inclination angle ($i < 82^\circ$), which in turn implies a lower
limit for the masses of the components ($M_1>
0.3646\;\mathrm{M_\odot}$ and $M_2> 0.3482\;\mathrm{M_\odot}$).

As expected for stars in such a close orbit, their rotational periods
are synchronised to the orbital period. This is observed as a
modulation in the V-band light curve with the same period as the
orbit. Additionally, the amplitude of this modulation is clearly seen
to change with time in longer timescales, which we believe is due to
differences in the area of the stars covered by the active regions.

Therefore, by studying the evolution of the amplitude of the
rotational modulation we found that the system exhibits a roughly
periodic behaviour of 2.2 years (or $\sim$ 800 days). The same period
was found in the mean magnitude of the system and in the flux of the
Ca II K line, although the Ca flux variations occur 140 days ahead of
the photometric ones. This behaviour was observed previously in other
single stars and binary systems, for which the lag between temperature
variations and line fluxes was reported to anticorrelate with
effective temperature~\citep{gray96b}. The measured lag for this
system fits roughly in that trend, indicating that the same process
may be responsible for the phase shift between observables.

The agreement between the behaviour of the three observables is
remarkable, because of the different nature of the observations, and
the different instruments and sites in which they were
obtained. Additionally, we performed a non linear fit to the total Ca
II K flux using a sinusoidal function, and found the least-square
estimation for the period is $P=763$ days. If confirmed, this would be
the second detection of an activity cycle in an M star, the first
being a recently reported cycle of around 450 days in Proxima
Centauri~\citep{proxima}. It is interesting to compare briefly both
cases in relation with the 2.5-years lower limit for the period of
activity cycles in solar stars given by~\citet{baliunas95}. Although
slightly shorter, the period of the Gl 375 system is consistent with
this limit, which is reasonable, since Gl 375 A and Gl 375 B are dM3.5
stars, and should therefore have a radiative core surrounded by a
convective outer envelope. Therefore, in the Gl 375 system the
standard dynamo could in principle explain the cyclic behaviour
observed, as in the earlier stars included in the study by
\citet{baliunas95}. On the other hand, the period of Proxima Centauri
is well below the 2.5-years cutoff. Since Proxima Centauri is a dM5.5
star, and should therefore be completely convective, a possible
interpretation of this fact is that different mechanisms are
responsible for the production of the activity cycle in Gl 375 and in
Proxima Centauri. Further observations of these systems, as well as
the discovery of activity cycles in other M stars should provide a
solid observational base to develop a model capable of explaining the
observations.

Another interesting result of this work is that the activity of Gl 375
A and Gl 375 B, as measured in the flux of the Ca II K lines, are in
phase. The levels of chromospheric emission of both components exhibit
an excellent correlation, which implies that a magnetic connection
exists between both components, as is shown in numerical simulations
by other authors. Due to its vicinity and relative brightness this
system presents an interesting opportunity to further study this type
of interaction.

Finally, since the chromospheric emission is larger when the system is
brighter, and this coincides with period of maximum amplitude of the
modulation, we conclude that the photometry of these stars is
dominated by bright active regions rather than dark spots and that the
filling factor is below 50\%, very much like in the Sun.

\begin{acknowledgements}
The CCD and data acquisition system at CASLEO has been partly financed
by R. M. Rich through U.S. NSF grant AST-90-15827. This research has
made use of the SIMBAD database, operated at CDS, Strasbourg,
France. We would like to thank the CASLEO staff and thankfully
acknowledge the comments and suggestions of the referee, which helped
us to improve our original manuscript.
\end{acknowledgements}

%\bibliographystyle{aa}
%\bibliography{/home/rodrigo/LaTeX/biblio}

\end{document}